\documentstyle[12pt,titlepage]{article}

\newcommand {\h} {$h^{-1} \, Mpc \,$}
\def\ref{\par\noindent\hangindent=1cm}
\begin{document}
\title{}
\author{}
\date{}
\vspace{4cm}
\Huge
\begin{center}
{\bf Structures \protect\\
in Galaxy Clusters}
\end{center}
\vspace{2cm}
\normalsize
\begin{center}
{\em E. Escalera$^{1}$,A. Biviano$^{2}$, M. Girardi$^{3}$,
G. Giuricin$^{3,4}$, F. Mardirossian$^{3,4}$, A. Mazure$^{1}$},
and {\em M. Mezzetti$^{3,4}$}
\end{center}
\vspace{0.5cm}
(1) Universit\'e de Montpellier II / C.N.R.S.,
place Eug\`ene Bataillon, 34095 - Montpellier, France \\
(2) Institut d'Astrophysique de Paris,
98bis, bd. Arago, 75014 - Paris, France \\
(3) SISSA, strada Costiera 11, 34014 - Trieste, Italy \\
(4) Dipartimento di Astronomia,
Universit\`{a} degli Studi di Trieste, via Tiepolo 11, 34131 - Trieste, Italy
\\

\pagebreak
\renewcommand{\thesection}{\arabic{section}}
\renewcommand{\thesubsection}{\thesection.\arabic{subsection}}
\baselineskip 12pt

\section*{Abstract}
The analysis of the presence of substructures  in 16 well-sampled clusters of
galaxies suggests a stimulating hypothesis: Clusters could be classified as
unimodal or bimodal, on the basis of to the sub-clump distribution in the {\em
3-D} space of positions and velocities. The dynamic study of these clusters
shows that their fundamental characteristics, in particular the virial masses,
are not severely biased by the presence of subclustering if the system
considered is bound.

\vspace{1cm} {\em Subject headings:} \ galaxies:
clustering

\pagebreak
\section{Introduction}
The presence or absence of substructures in galaxy clusters is
very important in cluster evolution, as well as in the cosmological
distribution of galaxy systems. In fact, subclustering may give constraints
on the epoch of cluster formation, their present evolutionary status, the
existence and distribution of dark matter, the morphology of galaxy members,
etc. (see, e.g., Cavaliere et al. 1986; West, Oemler, \& Dekel 1988; Fitchett
1988a). Moreover, the frequency of substructures in clusters offers the
possibility of constraining the density parameter $\Omega$; in fact,
structures (and substructures) should collapse earlier in low-density universes
than in high-density universes (see, e.g., Richstone, Loeb, \& Turner
1992; Lacey \& Cole 1993; Kauffmann \& White 1993; Mamon 1993).

Both Abell and Zwicky (see, e.g., Abell 1964; Abell, Neymann, \& Scott 1969;
Zwicky 1957; Zwicky et al. 1961-1968) considered the hypothesis that clusters
contain substructures. Subsequently, contour maps in optical and X-ray domains
(see, e.g., Baier 1979,1980; Jones et al. 1979; Geller \& Beers 1982; Ulmer,
Cruddace, \& Kowalski 1985; Forman \& Jones 1990) showed evidence of cluster
substructures. West, Oemler, \& Dekel (1988) contributed to the study of the
problem of subclustering with alternative methods, which consider different
kinds of spatial substructures. Other methods are the {\em Lee-Method}, which
is based on maximum-likelihood (see Lee 1979; Fitchett \& Webster 1987;
Fitchett 1988), the {\em Delta Statistics} developed by Dressler \& Shectman
(1988), and the method devised by Gurzadyan, Harutyunyan,\& Kocharyan (1991).
These three methods are able to use both galaxy positions and velocity field,
and they provide the statistical significance of the substructures detected in
the cluster.

The recent literature contains numerous papers with observational evidence of
cluster subclustering both in the optical and in the X-ray domains (see, e.g.,
Bahcall 1973; Quintana 1979; Forman et al. 1981; Henry et al. 1981; Forman \&
Jones 1982; Geller \& Beers 1982; Ulmer \& Cruddace 1882; Bothun et al. 1983;
Huchra, Davis, \& Latham 1983; Forman \& Jones 1984; Lucey, Currie, \& Dickens
1986; Bingelli et al. 1987; Fitchett \& Webster 1987; Colles \& Hewett 1988;
Mellier et al. 1988; Mushotzky 1988; Forman \& Jones 1990; West \& Bothun 1990;
Beers et al. 1991; Briel, Henry, \& Boehringer 1991;
Rhee, van Harleem, \& Katgert 1991; Jones \& Forman 1992;
Escalera et al. 1993a; White, Briel, \& Henry 1993).

The observational situation is combined with a rich theoretical background.
N-body codes have been widely used to describe cluster evolution, as well as
the growing and the dumping of the subclustering (see, e.g., White 1976,
Cavaliere et al. 1986; Fitchett 1988a; West, Oemler, \& Dekel 1988; Cavaliere
\& Colafrancesco 1990; Cavaliere, Colafrancesco \& Menci 1991; Schindler \&
Boehringer 1992). The possible presence of bimodal structures has been
considered by Cavaliere et al. (1986).

The results of theoretical simulations and observational tests have been often
debated and at present do not provide a final, coherent picture of
subclustering in galaxy clusters.

In the present paper we use {\em Weighted Wavelet Analysis} to look for
substructures in 16 well-sampled galaxy clusters. We plan to look for evidence
of subclustering from a dynamic point of view, inasmuch as we combine the
spatial distributions of the galaxies with their velocities.

\section{Data}
For the detection of substructures, only well-sampled, complete (or
quasi-complete) clusters of galaxies are suitable for study. The 16 systems we
studied are nearby clusters with a limited range of redshifts (chosen to
minimize possible evolutionary effects). For some clusters, several samples
were available, at different levels of completeness in velocity $-$ 85\%, 95\%,
and 100\%. The completeness in velocity was measured by the ratio of the number
of cluster members with redshift over the total number of cluster galaxies
present in the field, up to a given magnitude. We then removed those samples
containing fewer than some 30 galaxies, since for poorly populated structures
it is not possible to obtain reliable results on their dynamics.

In Table 1,  Column (1) lists the names of clusters; Column (2) the richness
classes $R$, mainly taken from Abell, Corwin, \& Olowin (1989); Columns (3) and
(4) give the relevant references for velocities and magnitudes, respectively;
Columns (5) and (6) give the coordinates $\alpha_0$ and $\delta_0$,
respectively, of the origin of the map (bottom-right corner); Columns (7) and
(8) show the extension of the map along the X and Y axies, respectively, given
in arcmin.

The criteria with which the cluster members have been selected are the same as
in Girardi et al. (1993). A maximum  aperture of 3 \h  picks out our clusters
(we use $H_{0}= 100 \, h \, km \, s^{-1} \, Mpc^{-1}$ throughout). Our virial
masses have been obtained by means of robust estimates of the velocity
dispersions (see, e.g., Girardi et al. 1993)

\section{Wavelet Analysis}
The Wavelet Transform is the convolution between the signal (i.e., the angular
positions of galaxies) and an analysing wavelet function. As a result of such a
convolution, the signal is transformed into a set of "coefficients", the {\it
wavelet coefficients}, which contain all the information we need for the
analysis of cluster structures.

The particularities of a wavelet versus other more "classical" functions are
the following:

(i) A wavelet is naturally bound, and invariant with translation: so it is
particularly suitable for a {\em local} analysis (Time-Frequency, or, for the
present work, Position-Size);

(ii) It is also invariant with dilatation, allowing to a {\em multi-scale}
analysis.

So the main analysis consists in performing the Wavelet Transform at each point
of the signal, using a full set of scales: thus no structure in the signal will
escape detection, whatever its location and size.

A wavelet is a function that obeys some concrete mathematical conditions
(continuity, differentiability, \ldots etc.; see Escalera et al. 1993a for
details). It always has a null mean value. Thus a constant signal will lead to
null coefficients. In this work we use the 2$D$ radial wavelet, called Mexican
Hat, which comes from  a second derivation of a gaussian: as a consequence, a
constant gradient will also produce null coefficients. This is particularly
useful for detecting small structures inserted in the gradient of a big one.

If now the signal presents some discontinuities, the wavelet coefficients will
react accordingly, producing a local maximum. The stronger the discontinuity
correlated to the analysing wavelet (size/scale), the stronger the value of the
corresponding local maximum.

The data forma map of the analysed cluster with galaxies plotted as points. So
our signal is a 2$D$ distribution of $\delta$-functions, and the structures are
then defined as a local excess of density. Considering the above working of the
transform, we understand that areas of uniform distribution will give null
coefficients, while substructures of a given size will produce high local
values of the wavelet coefficients (hereafter called local maxima). In all
cases the edge effects are taken into account, since wavelets are very
sensitive to density contrasts (see also Escalera \& Mazure 1992). In this work
the densities reach zero value before the edge of the field, so the edge
effects can be easily removed by analysing areas greater than the cluster, with
null areas outside the limits of the true field.

Therefore, as mentioned above, the main analysis consists in performing the
wavelet transform with a set of different scales.

For a given scale $s$, the Mexican Hat explores, at each pixel, a circular area
within a radius roughly equal to 4$s$. The largest scale that we consider
is $s$ = 0.25 $R_f$ (where $R_f$ is the radius of the analysed field), so the
main cluster should be detected as a single structure. On the other hand,
the lower limit of the wavelet scales is estimated from the mean inter-particle
distance : the corresponding explored area contains only one galaxy. Thus, the
lowest value of the wavelet scale $s$ is given by the relation $(4 s)^2 = R_f^2
/ N_g$ (where $N_g$ is the number of galaxies considered). The usual values of
$N_g$ generally mean that the smallest scale is equal to 0.05$R_f$.

In this paper, each analysed field corresponds to a map of $256 \times 256$
pixels. The wavelet scales used, given in pixel units, are $s$ = $64$, $32$,
$24$, and $16$; for all the analysed samples, the largest and smallest scales
used are, respectively, 64 and 16 pixels. The field analysed with the largest
scale will produce a wavelet image showing generally a single central
structure, hereafter called A-structure, which corresponds to the detection of
the whole cluster. If the scale is decreased, many other events can be
observed. The central A-structure may remain regular at every scale, or may
split into substructures. Minor substructures may also appear in any region of
the cluster field.

The results of such an analysis, as extracted from the local maxima, are the
following :

(a) a given structure is located in the field by the position of the local
maximum of the wavelet coefficients;

(b) its size is estimated from the scale which best detects the structure, i.e.
the scale producing the greatest value of the local maximum of the wavelet
coefficients; in the following, such an optimal wavelet scale is called
W-scale;

(c) the statistical significance is derived directly from this local maximum,
and gives the individual probability of existence of the concerned structure
(via random simulations, see below).

Thus a given structure, even if detected at many scales, is only defined in
terms of {\em position-dimension-probability} by its corresponding W-scale. In
this way we detect all the structures lying on the map, whatever their
location and size may be.

As a fundamental point of our analysis, we note that the above results are
individually obtained for each structure present in the signal. We may also
recall that in this paper we perform the {\it Weighted} Wavelet Analysis. This
method gives weights to the galaxies by considering the velocity data, in a way
that takes into account the correlations that exist in physical substructures.
As a weight we used the $\delta$ parameter introduced by Dressler \& Shectman
(1988); see also Escalera \& Mazure (1992) and Escalera (1992) for further
details on the weighting procedure. Thus the detected structures contain 3$D$
information. All these individual 3$D$ results make possible an interpretation
of the meaning of the detected structures from a dynamic point of view.

The wavelet method quantifies the statistical validity of the detection of
substructures by using sets of Monte Carlo simulations of the analysed
clusters. Such simulations are performed in order to remove the real
correlations (in positions {\it and} velocities) that may exist in the analysed
cluster. Thus, a given local maximum of the wavelet coefficients, which
corresponds to a true structure, should not be reproduced in the simulations.
In fact, we check how many times a structure similar to that detected in the
real cluster appears in these random simulations. The natural parameter which
makes this comparison possible is the value of the local maximum of the wavelet
coefficients. This procedure gives us, for each structure, the probability of
its occurring by chance (see Escalera et al., 1993a, for details). As an
additional check, we coupled the analysis with the bootstrap resampling
technique (see, e.g., Efron \& Tibshirani 1986), which can also provide every
substructure with a significance level. In the present paper, almost all the
sub-clumps detected have significance levels $\geq$ 99\%. Probabilities from
both procedures are associated with the results.

\section{Different Morphologies}
When one considers the largest wavelet scales, two cases seem sufficient to
describe the morphology of our set of clusters:

- Case (U) : {\em Unimodal Clusters}. These clusters show a single dominant
central structure, which is located at the dynamic center of the cluster when a
detailed analysis is carried out. Case (U) concerns also clusters with a
structure that is single but considerably elongated, without any well-defined
features.

- Case (B) : {\em Bimodal Clusters}. These clusters show two dominant
structures, roughly equivalent in size and richness, with two well-separated
centers.

In this paper, due to the limited number of samples, we chose to include all
"elongated" cases in the (U) class, although these structures may mask true
bimodal clusters, e.g., ones closely aligned with the line of sight. Obviously
the study of a richer set of clusters might allow us to remove this ambiguity.

When we now explore smaller and smaller cluster-areas with decreasing scales,
two kinds of events enrich this morphological scenario:

- Event (sc) : {\em  structures in the core} . The central structures (dominant
at the largest scale) can remain single with decreasing scales or can split
into smaller substructures of group-sizes. This concerns also the bimodal
cases, when the splitting occurs in at least one of the two main components.

- Event (sf) : {\em  structures in the field } .  Small but statistically
relevant substructures appear in the whole cluster area when small scales are
used.

Both events $sc$ and $sf$ describe group-size features, each containing
some 10\% of the total cluster population.

The distinction between "core" and "field" areas is not induced only by the map
inspections since we took into account the 3-$D$ data, i.e. the positions and
velocities relative to the parameters of the central A-structure. In this way,
the $sf$ and $sc$ classifications are not subjective.

In Table 2 one can see the results of our analysis: 12 of our clusters are
unimodal and 4 are bimodal. In particular, Column (1) lists the names of the
clusters; Column (2) gives the sample used, with different velocity
completeness (** = 85\%, * = 90\%, no symbol = 100\%); Column (3) shows the
number of galaxy members considered in the analysis; in Column (4) the
morphological classification of substructures is coded with the symbols
explained above, $U$ and $B$ referring to the classification at the largest
wavelet scale, and $sc$ and $sf$ concerning the classification at smaller
scales.

Table 3 describes, cluster by cluster, the main structure (or the binary
structure) and the substructures detected. Substructures with fewer than 4
galaxies are not listed, since for such poorly populated structures it is not
possible to produce reliable conclusions from a dynamic analysis. Column (1)
lists the cluster name; Column (2) lists the samples used, with different
velocity completeness (the coding is the same as in Table 2); Column (3)
identifies the structure ($M$ = main cluster, $A$ = dominant central structure,
$A1, A2$ = two components in bimodal cases, $B, C,$... = substructures); Column
(4) gives the wavelet scales at which the structures were detected (the
so-called W-scale). Values are given in pixel units for a map of $256 \times
256$ pixels. Column (5) gives the number of galaxies belonging to the concerned
structures; Column (6) lists the probability that each structure is due to
random fluctuations (see \S 3); since our procedure for assigning this
probability checks the significance of subclustering by comparing the observed
distribution of galaxies to many random representations of a regular cluster
(single central structure), it is obvious that no probability can be assigned
to the A-structure; Column (7) gives the probabilities $P_{Lee}$, according to
the Lee-method; Column (8) gives the probabilities $P_{\Delta}$, according to
Dressler \& Schectman's (1988) method. In all cases (Columns (6),(7), and (8))
no substructure is considered as detected when the associated probability is $<
95\%$. There is a good agreement between our wavelet method and the others two
methods.

In order to make a compact presentation, we display only two wavelet images per
cluster although the multiplicity of structures in some particular cases would
require three or more images to show the W-scales of all clumps.

For each cluster we show the most complete sample, as listed in Table 2. In
each figure, right ascension grows from left to right. The coordinates of
origin (bottom-right corner of the map) as well as a map scaling correspondence
are given in Table 1. The wavelet images are superimposed over the cluster
galaxy plots. Solid lines represent the isopletes of the wavelet coefficients.
Labelled structures correspond to those discussed in Table 2. Non-labelled
features are those not statistically relevant or containing fewer than 4
galaxies.

Figures 1$A$ to 16$A$ show the wavelet images of our 16 clusters obtained by
using the large scales. In this way it is possible to identify unimodal ($U$)
and bimodal ($B$) clusters. Figures 1$B$ to 16$B$ show the wavelet images of
the same clusters obtained by using the small scales. Minor substructures
appear in some clusters, following the above-mentioned scenarios -- cases
($sc$)
and ($sf$). Notice that in general the central A-structure is still
seen at small scales, although the corresponding physical structure is only
defined by the largest scales.

\section{Substructure Dynamics}
The dynamics of the clusters and their corresponding substructures and, in
particular, the estimates of galaxy velocity dispersions and virial masses,
help to corroborate the morphological description of clusters given above.

Table 4 lists the most important physical parameters for the clusters and their
substructures. In particular, Column (1) lists the name of the clusters; Column
(2) lists the samples used, the degree of velocity completeness being coded as
in Table 2; Column (3) identifies the structure: $M$ = main cluster, $A$ =
dominant central structure, $A1,A2$ = two components in bimodal cases, $B,C,$
\ldots = substructures, and $M-BC$  = environment, {\em i.e.} main cluster with
substructures $B,C,$ \ldots removed. Each environment file well describes the
cluster as a whole, in particular when the removed substructures are very far
from the center, in terms of both position and velocity. Column (4) gives the
number $N_g$ of galaxies of the concerned structure or substructure; Column (5)
lists the mean velocity (biweight estimate), $V$, in $Km \, s^{-1}$; Column (6)
lists the robust velocity dispersion, $\sigma$, in $Km \, s^{-1}$; Column (7)
lists the virial mass, $M_v$, in units of $10^{14} \, M_{\odot}$; Column (8)
gives the virial radius, $R_v$, in $Mpc$. Obviously we have to put more weight
on the results obtained at a higher completeness level.

When reading Table 4. we may remark that, in some particular cases, a
substructure detected at a small scale is totally included in the dominant
structure detected at the largest scale (e.g., see $A$ and $B$ for the sample
A2670). On the other hand, in some cases the central $A$-structure detected at
the largest scale does not differ from the whole cluster (e.g., $A = M$ for the
sample S0463).

The robust velocity dispersion  of the whole cluster does not seem to be
strongly biased by the presence of subclustering in most of our 16 clusters.

Generally, the virial masses of the dominant structures (designated $A$, $A1$
and $A2$ in Tables) fairly closely approximate  the total virial mass of the
main cluster (designated $M$). In particular, the ratio of masses is rather
well related to the ratio of populations (numbers of galaxies involved in the
respective systems). This can be easily checked by comparing the masses and
populations of structures $M$ and $A$ for unimodal clusters, and structures $M$
and $A1+A2$ for bimodal clusters. The difference between the total mass and the
subclump masses turns out to be huge only when the detected substructures do
not belong to a single cluster, but are part of physically unbound systems.
This only happened once in our set, in the case of A0151, where two probably
unbound galaxy systems are seen in projection close to one another, looking
like a binary cluster. The assumption that a system is bound may be based on
the {\em Newton Gravitational Criterion} (see, e.g., Beers et al. 1982). So the
virial mass estimates of our clusters do not seem to be severely biased by the
presence of their substructures. Moreover, the masses of the small
substructures detected are generally one order of magnitude smaller than the
total mass of the cluster. This is consistent with the relative populations of
these small systems, which mostly represent, as mentioned above, some 10\% of
the total numbers of galaxies.

However, the virial analysis we have performed is but a partial test, and these
results should be checked both by extending the analysis to a larger sample of
clusters and by performing numerical simulations.

\subsection{Conclusions}
In our set of 16 clusters, subclustering seems to be able to classify cluster
structures quite naturally as $Unimodal$ and $Bimodal$.

We wish to consider this result as a stimulating hypothesis for cluster
classification. In fact, the possibility of classifying galaxy clusters by
means of their structure would allow us to combine their present dynamic
statuses, and therefore their evolutionary histories, with their morphological
characteristics. It may be interesting to note that our results are in
agreement with the cluster simulations by Cavaliere et al. (1986); these
Authors claim the existence of a clear bimodal configuration in about 30$\%$ of
their runs, and we have 4 bimodal systems in our 16 clusters. Salvador-Sol\'e,
Sanrom\`a, \& Gonz\'alez-Casado (1993) and Salvador-Sol\'e, Gonz\'alez-Casado,
\& Solanes (1993) find that at least $50 \%$ of apparently relaxed clusters
contain significant substructures, and we have only 3 clusters without
significant substructures. Moreover, both subclustering and binary structures
have been detected also in cluster X-Ray-maps (see, e.g., Forman et al. 1981;
Henry et al. 1981; Ulmer \& Cruddace 1882; Forman \& Jones 1984; Mushotzky
1988; Jones \& Forman 1992; White, Briel, \& Henry 1993), in particular
Mushotzky (1988) suggests that a significant fraction of all clusters ($\sim 25
\%$) shows X-Ray double images, and Jones \& Forman (1992) estimate that double
and complex structures are about $20 \%$.

The preliminary results concerning the dynamics of clusters and their
substructures seem to stress the robustness of the virial mass estimate and the
corresponding limited biases induced by subclustering. The virial masses of
substructures are closely related, in our clusters, to the number of galaxies
belonging to the substructures themselves. The typical masses of these
subtructures are often $\sim 10\%$ of the parent-cluster total mass. This is in
agreement with the merging-hierarchical-scenario suggested by Cavaliere,
Colafrancesco, \& Menci (1992) for galaxy clusters, where small galaxy systems
survive the merging processes inside the clusters.

The presence of these substructures and their frequency may also contribute to
the debated determination of the density parameter $\Omega$. In fact,
structures (and substructures) should collapse earlier in low-density universes
than in high-density universes. Therefore, these subclumps should not often be
detected today in clusters if $\Omega$ is small. However, this point is
sensitive to a correct estimate of the survival time of subtructures in a
collapsed cluster (see, e.g., Richstone, Loeb, \& Turner 1992; Lacey \& Cole
1993; Kauffmann \& White 1993; Mamon 1993).

Of course, the limited number of our clusters advise us to be wary and to look
for further confirmation of all our results in other optical cluster samples,
which should be more extended in number, richness classes, and redshift, as
soon as they become available.

\vspace{2cm}

We thank Alfonso Cavaliere and Nicola Menci for their useful discussions, and
the anonimous Referee for his/her suggestions. A.B. acknowledges partial
financial support from {\em Fondazione Angelo Della Riccia}. This work was
partially supported by the {\em Ministero per l'Universit\`a e per la Ricerca
scientifica e tecnologica}, and by the {\em Consiglio Nazionale delle Ricerche
(CNR-GNA)}.

\pagebreak \section*{References}

\ref Abell, G.O. 1964 A\&A, 8, 529.

\ref Abell, G.O., Corwin, H.G.Jr., \& Olowin, R.P. 1989, ApJS, 70, 1.

\ref Abell, G.O., Neymann, J., \& Scott, E.L. 1969, AJ, 69, 529.

\ref Bahcall, N.A. 1973, ApJ, 183, 783.

\ref Baier, F.W. 1979, Astron.Nachr., 300, 85.

\ref Baier, F.W. 1980, Astron.Nachr., 301, 17.

\ref Beers, T.C., Forman, W., Huchra, J.P., Jones, C., \& Gebhardt, K. 1991,
AJ, 102, 1581.

\ref Beers, T.C., Geller, M.J., \& Huchra, J.P. 1982, ApJ, 257, 23.

\ref Bingelli, B., Tammann, G.A., \& Sandage, A. 1987, AJ, 94, 251.

\ref Briel, U.G., Henry, J.P., Boehringer, H. 1991, in {\em Clusters and
Superclusters of Galaxies}, M.M. Colles, A.Babul, A.C. Edge, R.M. Johnstone, \&
S. Raychaudhury Eds.

\ref Cavaliere, A., \& Colafrancesco, S. 1990, in {\em Clusters of Galaxies},
W.R. Oegerle, M.J. Fitchett, \& L. Danly Eds., Cambridge University Press,
Cambridge.

\ref Cavaliere, A., Colafrancesco, S., \& Menci, N. 1991, in {Cluster and
Superclusters of Galaxies}, A.C. Fabian Ed., Series C: Mathematical and
Physical Sciences, Vol. 366, Kluwer Academic Publisher, Dordrecht.

\ref Cavaliere, A., Santangelo, P., Tarquini, G., \& Vittorio, N. 1986, ApJ,
305, 661.

\ref Colless, M., \& Hewett, P. 1987, $M.N.R.A.S.$, 224, 453.

\ref Dressler, A. 1980, ApJS, 42, 565.

\ref Dressler, A., \& Schectman, S.A. 1988, AJ, 95, 985.

\ref Efron, B., \& Tibshirani, R. 1986, Stat. Sci., 1, 54 .

\ref Escalera, E. 1992, PhD Thesis "{\em \'Etudes Dynamiques et Structurales
des Amas de Galaxies}", University of Paul Sabatier Press, Toulouse, France.

\ref Escalera, E., Biviano, A., Girardi, M., Giuricin, G., Mardirossian, F.,
Mazure, A., \& Mezzetti, M. 1993b, in preparation.

\ref Escalera, E., \& Mazure, A. 1992, ApJ, 388, 23.

\ref Escalera, E., Slezak, E., \& Mazure, A. Bijaoui, A. 1993a, A\&A, 264, 379.

\ref Fitchett, M.J. 1988a, in {The Minnesota Lectures on Clusters of Galaxies
and Large-Scale Structure}, J.M. Dickey Ed., Bringham Young University Press,
Provo.

\ref Fitchett, M.J., \& Webster, R.L. 1987, ApJ, 317, 653.

\ref Forman, W., Bechtold, J., Blair, W., Giacconi, R., Van Speybroek, L., \&
Jones, C. 1981, ApJ, 243, L133.

\ref Forman, W., \& Jones, C. 1982, ARA\&A, 20, 547.

\ref Forman, W., \& Jones, C. 1984, in {\em Clusters and Groups of Galaxies},
ed. F. Mardirossian, G. Giuricin, \& M. Mezzetti, Dordrecht:Reidel, p. 143.

\ref Forman, W., \& Jones, C. 1990, in {\em Cluster of Galaxies}, W.R. Oegerle,
M.J. Fitchett, \& L. Danly Eds., Cambridge University Press, Cambridge.

\ref Geller, M.J., \& Beers, T.C. 1982, PASP, 94, 421.

\ref Girardi, M., Biviano, A., Giuricin, G., Mardirossian, F., \& Mezzetti, M.
1993, ApJ, 404, 38.

\ref Gurzadyan, V.G., Harutyunyan, V.V., \& Kocharyan, A.A. 1991, in {\em 2nd
DAEC Meeting: The Distribution of Matter in the Universe}, G.A. Mamon, \& D.
Gerbal Eds., Observatoire de Paris, Paris.

\ref Henry, J., Henriksen, M., Charles, P., \& Thorstensen, J. 1981,
ApJ 243, L137.

\ref Huchra, J.P., Davis, R.J., \& Latham, D.W. 1983, in {\em Clusters and
Groups of Galaxies}, F. Mardirossian, G. Giuricin, \& M. Mezzetti Eds, Reidel
Pub. Co., Dordrecht.

\ref Jones, C. \& Forman, W. 1992, in {\em Clusters and Superclusters
of Galaxies}, A.C. Fabian Ed., Kluwer Pub. Co., Dordrecht.

\ref Jones, C., Mandel, E., Schwarz, J., Forman, W., Murray, S.S., \& Harnden,
F.R. 1979, ApJ, 234, L21.

\ref Kauffmann, G., \& White, S.D.M. 1993, MNRAS, 261, 921.

\ref Kent, S.M., \& Sargent, W.L.W. 1983, AJ, 87, 945

\ref Lacey, C.G., \& Cole, S. 1993, MNRAS, 262, 627.

\ref Lee, K.L. 1979, J.Am.Stat.Assoc., Vol.74, No.367, 708.

\ref Lucey, J.R., Currie, M.J., \& Dickens, R.J. 1986, $M.N.R.A.S.$, 221, 453.

\ref Mamon, G.A., 1993 to appear in {\em Gravitational Dynamics and
the N-body Problem}, F. Combes \& E. Athanassoula Eds.

\ref Mellier Y., Mathez, G., Mazure, A., Chauvineau, B., \& Proust, D. 1988,
A\&A, 199, 67.

\ref Mushotzky, R. 1988, in {\em Hot Thin Plasmas in Astrophysics}, R.
Pallavicini Ed., Kluwer Academic Publisher.

\ref Nilson, P. 1973, "Uppsala General Catalogue of Galaxies", Almquist \&
Wiksell, Stockholm, Sweden.

\ref Ostriker, E.C., Huchra, J.P., Geller, M.J., \& Kurtz, M.J. 1988, AJ, 96,
177.

\ref Quintana, H. 1979, AJ, 84, 15.

\ref Rhee, G.F.R.N., van Harleem, M.P., \& Katgert, P. 1991, A\&A, 246, 301.

\ref Richstone, D., Loeb, A., \& Turner, E.L. 1992, ApJ, 393, 477.

\ref Richter, O.G. 1987, A\&AS, 67, 237

\ref Richter, O.G. 1989, A\&AS, 77, 237

\ref Salvador-Sol\'e, E., Gonz\'alez-Casado, G., \& Solanes, J.M. 1993, ApJ,
410, 1.

\ref Salvador-Sol\'e, E., Sanrom\`a, M., \& Gonz\'alez-Casado, G.,  1993, ApJ,
402, 398.

\ref Sharples, R.M., Ellis, R.S., \& Gray, P.M. 1988, MNRAS, 231, 479.

\ref Schindler, S., \& Boehringer, H. 1992, preprint.

\ref Ulmer, M.P., Cruddace, R.G. 1982, ApJ, 258, 434.

\ref Ulmer, M.P., Cruddace, R.G., \& Kowalski, M.P. 1985, ApJ, 290, 551.

\ref West, M.J., \& Bothun, G.D. 1990, ApJ, 350, 36.

\ref West, M.J., Oemler, A.Jr., \& Dekel, A. 1988, ApJ, 327, 1.

\ref White, S.D.M. 1976, MNRAS, 177, 717.

\ref White, S.D.M., Briel, U.G., \& Henry, J.P. 1993, MNRAS, 261, L8.

\ref Zwicky, F. 1957, {\em Morphological Astronomy}, Springer-Verlag, Berlin.

\ref Zwicky, F., Herzog, E., Wild, P., Karpowicz, M., \& Kowal, C. 1961-1968,
in {\em Catalog of Galaxies and of Clusters of Galaxies}, Vols. 1-6, California
Institute of Technology, Pasadena.

\pagebreak \section*{Captions to Figures}

\noindent {\bf Figs. 1$A$ to 16$A$:} Wavelet images of our 16 clusters obtained
by using the largest scales : 64 pixels for all clusters, except 32 pixels for
A2151 and A3716 (all for a $256 \times 256$ map). See \S 3 for details.
Unimodal ($U$) and bimodal ($B$) clusters are easily identified.

\noindent {\bf Figs. 1$B$ to 16$B$:} Wavelet images of the 16 samples obtained
by using the smallest scales : 16 pixels for all clusters (for a $256 \times
256$ map). See \S 3 for details. Minor substructures (labelled B, C, ...)
appear in some clusters, corresponding to the events $sc$ and $sf$ (see text).

\pagebreak \section*{Captions to Tables}

\noindent {\bf Table 1.} The Data-Sample. Column (1): Cluster name; Column (2):
Richness class; Columns (3) and (4): References to velocities and magnitudes,
respectively: [1]\ Proust et al. (1992); [2]\ Dressler (1980); [3]\ Kent \&
Sargent (1983); [4]\ Zwicky et al. (1961--1968); [5]\ Ostriker et al. (1988);
[6]\ Nilson (1973); [7]\ Dressler \& Schectman (1988b); [8]\ Richter (1987);
[9]\ Richter (1989); [10]\ Sharples, Ellis, \& Gray (1988). Columns (5) and
(6): Coordinates of the origin of the map (bottom-right corner), $\alpha_0$ and
$\delta_0$ respectively; Columns (7) and (8): Extension of the map along the X
and Y axies respectively, given in arcmin.

\noindent {\bf Table 2.} Cluster classification. Column (1): Cluster name;
Column (2): Sample used, with different velocity completeness (** = 85\%, * =
90\%, no symbol = 100\%); Column (3): Number of galaxy members; Column (4):
labels $U$ and $B$, respectively {\em Unimodal} and {\em Bimodal}, refer to the
classification at the largest wavelet scales; labels $sc$ and $sf$ refer to
events detected at smaller wavelet scales.

\noindent {\bf Table 3.} Cluster Structures and Substructures. Column (1):
Cluster name; Column (2): Sample used, with different velocity completeness (**
= 85\%, * = 90\%, no symbol = 100\%);  Column (3): Structures ($M$ = main
cluster, $A$ = dominant central structure, $A1, A2$ = two components in bimodal
cases, $B, C,$... = substructures); Column (4): Wavelet scales used for the
detection of the concerned structures, given in pixel units for a 256 $\times$
256 map. Column (5): Number of galaxies involved in the concerned structure or
substructure; Column (6): Wavelet probability of subclustering;
Column (7): Subclustering Lee probabilities; Column (8):
Subclustering Dressler \& Schectman probabilities.

\noindent {\bf Table 4.} Dynamical Quantities of Clusters and Cluster
Substructures. Column (1): Cluster name; Column (2): Sample used, with
different
velocity completeness (** = 85\%, * = 90\%, no symbol = 100\%);  Column (3):
Structures ($M$ = main cluster, $A$ = dominant central structure, $A1,A2$ = two
components in bimodal cases, $B,C,$... = substructures, $M-BC$... =
environment, {\em i.e.} main cluster with substructures $B,C$... removed.);
Column (4): Number of
galaxies involved in the concerned structure;
Column (5): Mean
velocity (biweight estimate), in $Km \, s^{-1}$; Column (6): Robust velocity
dispersion, in $Km \, s^{-1}$; Column (7): Virial mass, in units of $10^{14} \,
M_{\odot}$; Column (8): Virial radius, in $Mpc$.
\end{document}